\documentclass[aps,prb,twocolumn,superscriptaddress,showpa
cs,superbib,floats]{revtex4-1}
\usepackage{graphics}% Include figure files
\usepackage{epsfig}
\usepackage{amsmath}
\usepackage{bm}% bold math
\usepackage{hyperref}% add hypertext capabilities
\begin{document}
% You should use BibTeX and revtex.bst for references
%\bibliographystyle{apsrev}

% marks overfull lines with blackboxes
%\draft

% Use the \preprint command to place your local institutional report
% number on the title page in preprint mode.
% Multiple \preprint commands are allowed.
\preprint{}

%Title of paper
\title{
Probing the electron-phonon coupling in ozone-doped 
graphene by Raman spectroscopy 
}
% Optional argument for running titles on pages
%\title[]{}

% repeat the \author .. \affiliation  etc. as needed
% \email, \thanks, \homepage, \altaffiliation all apply to 
%the current
% author. Explanatory text should go in the []'s, actual 
%e-mail
% address or url should go in the {}'s for \email and \homepage.
% Please use the appropriate macro for the type of information

% \affiliation command applies to all authors since the last
% \affiliation command. The \affiliation command should follow the
% other information

\author{F. Alzina}
\email[E-mail: ]{francesc.alsina.icn@uab.cat}
\author{H. Tao}
\altaffiliation[Present address: ]{RIMNST, Jiao Tong 
University, 200240, Shanghai, China}
\author{J. Moser}
\author{Y. Garc\'ia}
\author{A. Bachtold}
\affiliation{CIN2(Institut Catal\`a de 
Nanotecnologia-Consejo Superior de Investigaciones 
Cient\'ificas), Campus UAB, 08193 Bellaterra, Spain}
\author{C. M. Sotomayor-Torres$^{1,}$}
\affiliation{CIN2(Institut Catal\`a de 
Nanotecnologia-Consejo Superior de Investigaciones 
Cient\'ificas), Campus UAB, 08193 Bellaterra, Spain} 
\affiliation{Instituci{\'o} Catalana de Recerca i Estudis 
Avan\c{c}ats (ICREA), 08010 Barcelona, Spain}
%\homepage[]{Your web page}
%\thanks{}
%\altaffiliation{}

%Collaboration name if desired (requires use of superscriptaddress
%option in \documentclass). \noaffiliation is required (may also be
%used with the \author command).
%\collaboration{}
%\noaffiliation

%\date{Submitted to PRB on \today}
\date{\today}

\begin{abstract}

We have investigated the effects of ozone treatment on 
graphene by Raman scattering. Sequential ozone 
short-exposure cycles resulted in increasing the $p$ 
doping levels as inferred from the blue shift of the 2$D$ 
and $G$ peak frequencies, without introducing significant 
disorder. The two-phonon 2$D$ and 2$D'$ Raman peak 
intensities show a significant decrease, while, on the 
contrary, the one-phonon G Raman peak intensity remains 
constant for the whole exposure process. The former 
reflects the dynamics of the photoexcited electrons 
(holes) and, specifically, the increase of the 
electron-electron scattering rate with doping. From the 
ratio of 2$D$ to 2$D$ intensities, which remains constant 
with doping, we could extract the ratio of electron-phonon 
coupling parameters. This ratio is found independent on 
the number of layers up to ten layers. Moreover, the rate 
of decrease of 2$D$ and 2$D'$ intensities with doping was 
found to slowdown inversely proportional to the number of 
graphene layers, revealing the increase of the 
electron-electron collision probability.

\end{abstract}

% insert suggested PACS numbers in braces on next line


%63.22.Rc Phonons in graphene  
%63.20.kd Phonon-electron interactions
%78.30.-j Infrared and Raman spectra  
%78.30.Na Fullerenes and related materials  
%78.67.Wj Optical properties of graphene  
 
%*********************************************************************

%\maketitle must follow title, authors, abstract and \pacs
\maketitle

%*********************************************************************
% body of paper here - Use proper section commands

\section{Introduction}

Graphene linear carrier dispersion in the vicinity of two 
inequivalent points ($\textbf{K}$, $\textbf{K'}$) of the 
Brillouin zone creates the conditions for the occurrence 
of unusual effects on the dynamics of both electrons 
(holes) and phonons, which are related to the 
electron-phonon ($e$-$ph$) 
interaction.~\cite{Ando06,Lazzeri06,Samso07}

In graphene, doping can be tuned by means of the field 
effect, i.e., electric charge induced by capacitive 
coupling.~\cite{Novoselov04} Moreover, being a system 
entirely exposed to its environment, modification of the 
carrier concentration in graphene can follow from 
molecules adsorbed on the surface by charge 
transfer.~\cite{ISI:000264637800018} The control of the 
carrier concentration allows the study of electron-phonon 
coupling (EPC) effects since the $e$-$ph$ interaction is 
directly modified by changing the Fermi energy level. 
Raman scattering measurements in field-effect devices 
showed the dependence of the $G$ peak position and 
linewidth with doping, where $G$ is the one-phonon mode at 
the $\bm{\Gamma}$ point, unveiling tunable optical phonon 
anomalies.~\cite{Pisana07,Yan07,Das08,Yan08,Das09} The 
possibility to monitor doping-oriented studies in graphene 
by Raman spectroscopy has provided the basis for a large 
range of 
application.~\cite{ISI:000251678700021,ISI:000260888600008,
ISI:000262519100065,shi:115402}

Besides the effects of doping on the frequency and 
linewidth of the Raman $G$ peak, the intensity of the 2$D$ 
two-phonon signature, or the ratio of peak intensities 
$I(2D)/I(G)$, was found to decrease with increasing doping 
\cite{Das08} and used as a tool to qualitatively establish 
the presence of charged 
impurities.~\cite{casiraghi:apl07,ISI:000264535200013} 
More recently, an understanding of how the two-phonon 
Raman peaks intensity depends on doping has been provided 
based on fully resonant processes~\cite{Basko07,Basko08} 
and their dependence on the $e$-$ph$ and electron-electron 
($e$-$e$) collision rates established. In 
Refs.~\onlinecite{Basko09,casiraghi:233407}, the $e$-$ph$ 
scattering rate was not entirely obtained from experiments 
as the analysis of the $2D$ peak intensity requires the 
calculation of the $e$-$e$ scattering rate.

Ozone treatment is considered as a promising route to 
enhance the otherwise weak chemical reactivity of 
graphitic structures.~\cite{Tracz09,Lee09,Yim09} The 
conductivity of carbon nanotubes (CNTs) was found to 
increase at low ozone dose or exposure 
time.~\cite{picozzi:1466,Ma2009158} The proposed mechanism 
for this increase was the ozone adsorption on the CNT 
surface, which induces charge transfer effects. At high 
ozone dose (or exposure time), the generation of 
structural modifications and defects seems to be the cause 
of a reduced conductivity.~\cite{Ma2009158} In this paper 
we investigate the effects of sequential ozone treatment 
cycles on graphene flakes by analyzing the Raman spectrum. 
Our studies show that graphene displays similar changes 
with ozone as those reported in CNTs, i.e., $p$-type 
doping without introducing significant disorder at low 
exposure, and different degrees of bond disruption and 
surface etching at high exposure. Here, we restrict our 
study to low exposure conditions and a full account on 
graphene oxidation by ozone will be published 
elsewhere.~\cite{Tao} Raman spectroscopy tracks the 
process of gradual $p$ doping of the samples, as concluded 
from both the position and the intensity of the Raman 
peaks. From the latter we could determine the EPC for the 
phonon modes near the $\textbf{K}$ point, and to monitor 
the $e$-$e$ scattering contribution with increasing charge 
concentration as well as with the number of graphene 
layers. We found a good correlation between the rate of 
the intensity decrease upon doping and the number of 
graphene layers. 
\section{Samples and experimental method}
Graphene sheets, prepared by micromechanical cleavage of 
highly-ordered pyrolytic graphite 
(HOPG)~\cite{Novoselov04}, were deposited on Si wafers 
with 300 nm thick thermal silicon oxide. Ozone treatment 
of the samples was performed at room temperature with a 
Novascan UV-Ozone cleaning system. The morphology of the 
graphene sheets was studied using a Nanoscope AFM in 
tapping mode. Micro-Raman measurements were carried out at 
room temperature in backscattering geometry using a T64000 
Jobin-Yvon spectrometer with a cooled 
charge-coupled-device detector. In micro-Raman 
measurements the light was focused to a spot with diameter 
of about 1 $\mu$m through a 100x objective. The 514.5 nm 
emission line of an Ar$^+$ laser was used for excitation 
with a typical power of only 120 $\mu$W, in order to 
prevent structural damage of the sample surface by the 
laser irradiation.~\cite{Klitzing09} Raman peak lineshape 
and visible light absorption were used to determine the 
number of layers.~\cite{note1}

\begin{figure}
%[tb]
\centerline{\epsfig{file=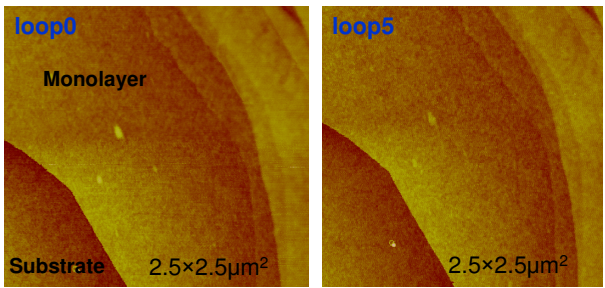,width=8.0 cm,clip}}
\caption{AFM images obtained from the same graphene sheet 
before ozone treatment (loop 0) and after five exposure 
cycles to ozone (loop 5).}
\label{AFM}
\end{figure}

Graphene samples were placed on the ozone cleaner for 
cycles of fixed duration. After each exposure, samples 
were analyzed morphologically by AFM and by Raman 
spectroscopy. The exposure time of 2.5 min during each 
cycle ensures the graphene surface quality is preserved. 
In Fig.~\ref{AFM} AFM images taken from the same sample 
region before and after five exposure cycles 
[Fig.~\ref{AFM}(loop 0) and (loop 5), respectively] reveal 
smooth surfaces. Here, loop 0 refers to no exposure to 
ozone while loop $n$ refers to a sample exposed to $n$ 
consecutive loops. This observation is supported by Raman 
scattering measurements showing that the structural 
disorder-related $D$ peak is barely detected, which 
indicates a low amount of damage of the graphene surface 
after being exposed to ozone.
%
%
%**********************************************************
% Figure 1 AFM images
%
%**********************************************************
%
\section{Results and Discussion}
The Raman spectrum of graphitic structures shows common 
features, i.e., the main one-phonon $G$ peak [cf. 
Fig.~\ref{Ramanserie}(a)], and defect-assisted one-phonon 
$D$ and $D'$ peaks, the expected frequencies of which are 
indicated by arrows in 
Fig.~\ref{Ramanserie}(a)
.~\cite{Tuinstra70,*Nemanich79,*Kawashima95,*Ferrari00, 
*Thomsen00} The $G$ and $D'$ peaks correspond to phonons 
at and near the Brillouin $\bm{\Gamma}$ point ($E_g$ 
mode), respectively. The $D$ peak comes from phonons near 
the $\bm{K}$ points ($A_1$ mode). Both $D$ and $D'$ peaks, 
which are evidence of inter- and intra-valley double 
resonance processes, respectively, require defect 
scattering for their activation. Thus, these peaks are 
only detected when carbon planes present structural 
imperfections.~\cite{Reich04,*Ferrari07} Notice the 
undetectable presence of the latter modes in the spectrum 
of the pristine sample [cf. Fig.~\ref{Ramanserie}(a)(loop 
0)] indicating high structural order.

The peaks denoted as 2$D$ and 2$D'$ in 
and $D'$ peaks, respectively, involving two-phonon 
processes with opposite wavectors, which do not require 
the presence of defects for their activation. The 
strongest and featureless 2$D$ peak in monolayer graphene 
evolves to a structured lineshape as the number of layers 
increases, revealing the electronic band structure, which, 
in turn, depends on the number of stacked 
layers.~\cite{Charlier08}
\subsection{Second-order Raman peaks intensity}
\begin{figure}
%[tb]
\centerline{\epsfig{file=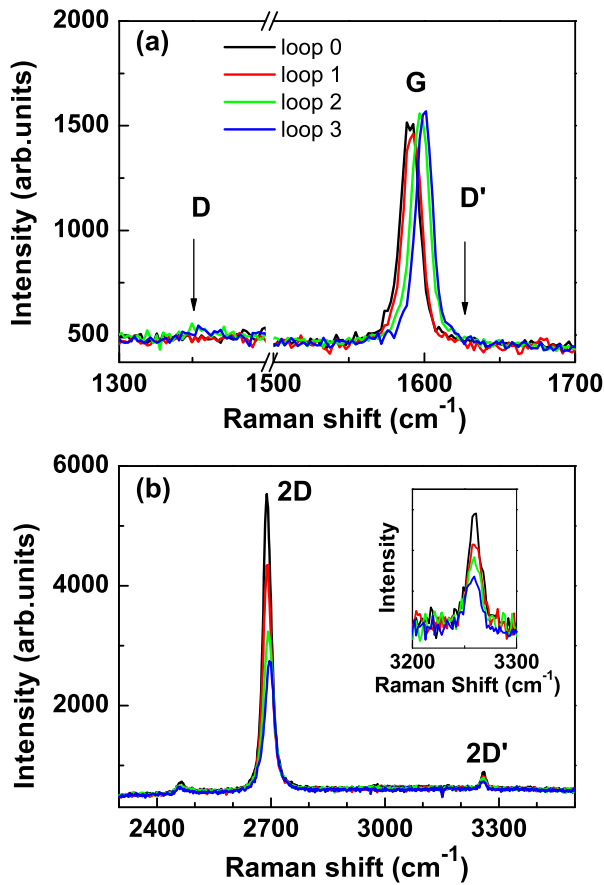,width=8.0 cm, clip}}
\caption{(a) First- and (b) second-order Raman spectra of 
single-layer graphene taken from the pristine sample (loop 
0) and after each exposure to ozone up to three cycles 
(loop 1 to loop 3). The inset shows the enlarged spectra 
in the spectral range of the $2D'$ Raman peak.}
\label{Ramanserie}
\end{figure}

With increasing ozone exposure, the two-phonon 2$D$ and 
2$D'$ Raman peak intensities show a significant decrease 
[see Fig.~\ref{Ramanserie}(b) and inset to 
Fig.~\ref{Ramanserie}(b)] while, in contrast, the 
one-phonon $G$ Raman peak intensity remains unchanged with 
ozone doping.
%
%
%**********************************************************
% Figure 2 Raman Spectra ozone treatment monolayer loop0 
%to loop 3
%
%**********************************************************
%

The activation of 2$D$ and 2$D'$ peaks involves four-step 
processes where all the states are real, and require 
energy and momentum conservation at every elementary step, 
which means that both two-phonon Raman processes are fully 
resonant. As a consequence, two-phonon Raman spectroscopy 
is sensitive to the dynamics of the photoexcited 
electron-hole ($e$-$h$) pairs, i.e., other than $e$-$ph$ 
inelastic-scattering processes they can undergo, for 
example, $e$-$e$ collisions.~\cite{Basko07} Assuming that 
these two processes are the main scattering mechanisms, 
the integrated intensities over the full linewidth of the 
2$D$ and 2$D'$ Raman peaks, which represent the 
probabilities of the respective Raman processes, can be 
expressed as:~\cite{Basko08}
\begin{subequations}
\label{eq:Intensities}
\begin{eqnarray}
 A(2D) = C (\gamma_K/\gamma)^2
 \label{subeq:2}
\end{eqnarray}
\begin{equation}
 A(2D') = C' (\gamma_\Gamma / \gamma)^{2},
\label{subeq:1}
\end{equation}
\end{subequations}
where $C$ and $C'$ are constants and 2$\gamma$ denotes the 
inelastic scattering rate of the photoexcited $e$-$h$ pair 
written as: 
\begin{equation}
 \gamma = \gamma_{e-ph}+ \gamma_{e-e},
\end{equation}
and the phonon emission rate  $\gamma_{e-ph}$ includes 
phonons near $\bm{\Gamma}$ and $\textbf{K}$,
\begin{equation}
 \gamma_{e-ph} = \gamma_{\Gamma}+ \gamma_{K}.
\end{equation}

Since the $e$-$e$ scattering is dependent on carrier 
density, both 2$D$ and 2$D'$ intensities are sensitive to 
doping level.~\cite{Basko09} The $e$-$e$ scattering rate, 
2$\gamma_{e-e}$, was found \cite{Basko09} to be 
proportional to the Fermi energy, $E_{F}$, and up to first 
order in $E_{F}$ it is expressed as:
\begin{equation}
\gamma_{e-e} = \left|E_{F} \right| f ,
\label{eerate}
\end{equation}
where the proportionality coefficient $f$ depends on the 
Coulomb coupling constant. The intensity of the $G$ peak 
should not depend on doping, since the given Raman process 
is in off-resonance condition.~\cite{Basko08} Therefore, 
the Raman peak intensity variation shown in 
Fig.~\ref{Ramanserie}, i.e., decrease of two-phonon Raman 
peaks intensity and unchanged $G$ peak intensity, could be 
attributed to a change of the carrier concentration due to 
ozone treatment. 
\subsection{Frequency shift of Raman bands}
The above interpretation is reinforced by monitoring the 
changes in frequency of the Raman spectrum feature after 
each ozone cycle, given the extensively reported 
dependence of the Raman modes frequency on doping in 
graphene. Charge-transfer modification of the 
chemical-bond induces 
variations of bond lengths (stiffening/softening), 
reflected in the variation of the phonon frequency. It has 
been shown that in graphene this effect alone does not 
explain the behavior of the Raman peaks frequency with 
doping, and therefore effects arising from the suppression 
of the Kohn anomaly at $\bm{\Gamma}$ and $\textbf{K}$ 
points must be invoked.~\cite{Lazzeri06} 

The $G$ peak frequency increases for both $n$ and $p$ 
doping, due to the nonadiabatic removal of the Kohn 
anomaly from the $\bm{\Gamma}$ 
point.~\cite{Pisana07,Yan07} Anomalous phonon softening, 
which is seen at low temperature but is smeared at 300 K, 
reflects a resonant $e$-$ph$ coupling effect when the 
$e$-$h$ energy gap is smaller than the phonon 
energy.~\cite{Yan08} Moreover, when the $e$-$h$ energy gap 
reaches a value higher than the phonon energy a sharp 
linewidth reduction occurs as the phonon decay process 
into $e$-$h$ pairs suddenly ceases.~\cite{Pisana07,Yan07}

The 2$D$ peak shows a different dependence on doping, 
which helps discerning between $n$- and $p$-type 
doping.~\cite{Das08,Das09} For electron doping, the 2$D$ 
peak frequency does not change much until a high electron 
concentration is reached, showing a softening for further 
increase. For hole doping, the frequency of the 2$D$ peak 
increases at a rate higher than solely expected from 
variations of lattice 
spacing.~\cite{Das08,Das09,saha:201404} Although phonon 
modes contributing to 2$D$ peak are away from $\textbf K$ 
points, the effects of the Kohn anomaly are not negligible 
due to the strong $e$-$ph$ coupling. As the doping level 
increases, the Kohn anomaly close to $\textbf K$ points is 
smeared out contributing to the stiffening of the 2$D$ 
peak. In contrast, the influence of the Kohn anomaly at 
$\bm \Gamma$ becomes weaker as the phonon wavevector 
departs from this point. Therefore, the 2$D'$ peak 
frequency is expected to be almost insensitive to changes 
in doping level. 

Finally, charge concentration changes are not the only 
possible source of phonon spectrum variation. Strain 
effects have already been measured in the Raman spectrum 
of graphene.~\cite{Mohiuddin09} The splitting of the $G$ 
band into two components displays a shift with applied 
uniaxial strain at a rate of 11 and 32 cm$^{-1}$/\%. The 
2$D$ and 2$D'$ bands do not split and they show a shift 
amounting to 64 and 35 cm$^{-1}$/\%, respectively.

\begin{figure}
\centerline{\epsfig{file=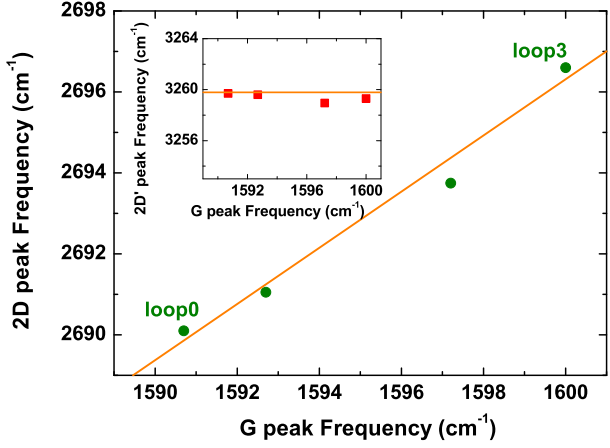,width=8.0 cm}}
\caption{Frequency shifts of the Raman 2$D$ vs $G$ peak 
obtained after each ozone exposure cycle. The inset shows 
the evolution of the Raman 2$D'$ vs $G$ peak frequency.}
\label{Freqshift}
\end{figure}

Figure \ref{Freqshift} displays the Raman $G$ and 2$D$ 
peak frequencies showing a blue shift with increasing 
exposure to ozone. The Raman 2$D'$ peak position, on the 
contrary, displays no variation within the experimental 
resolution (see inset to Fig.~\ref{Freqshift}). The larger 
frequency variation of the $G$ peak compared to the $2D$ 
peak, together with a constant $2D'$ peak frequency, rules 
out the effect of strain. Therefore, we conclude that the 
graphene surface increases its $p$ doping level with each 
ozone exposure cycle. The present method to change the 
amount of doping does not require additional processing to 
fabricate contacts, which can affect the crystal quality 
and the homogeneity of the properties of the graphene 
flakes. Comparing the measured phonons frequency shift in 
the present study with those found in the literature, we 
obtain the overall change of the carrier concentration to 
be $\Delta p_{tot}\approx 5 \cdot 10^{12}$ cm$^{-2}$. The 
Raman spectrum after four ozone exposure cycles (not 
displayed) shows no further changes, neither in the 
two-phonon peak intensities nor in the peak position of 
the $G$ and $2D$ features, indicating that the charge 
concentration, i.e., the ozone adsorption, reached a 
constant value.

%
%**********************************************************
% Figure 3 2D and 2D' mode frequency vs G mode frequency
%
%**********************************************************
%
%
%
\subsection{Electron-phonon coupling parameters ratio}
\begin{figure}
\centerline{\epsfig{file=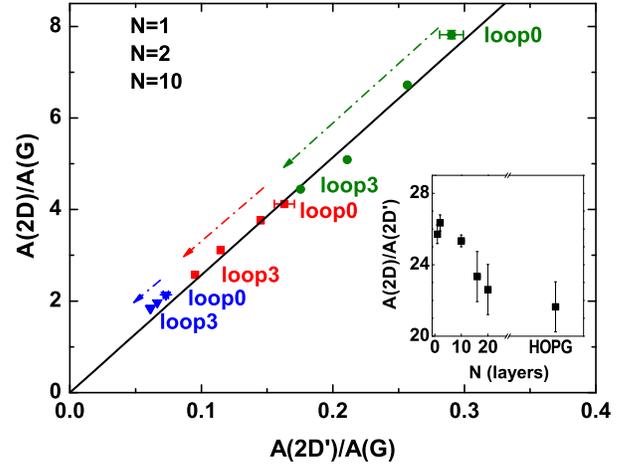,width=8.0 cm, clip}}
\caption{Experimental A(2$D$)/A($G$) vs A(2$D'$)/A($G$) 
Raman peaks area ratio from single-layer graphene (solid 
circles), double-layer graphene (solid squares) and 
ten-layer graphene (solid triangles) spectra measured 
before ozone treatment and after each ozone exposure 
cycle. The inset shows the A(2$D'$)/A($2D$) ratio as a 
function of the number of graphene layers, $N$.}
\label{Ratio}
\end{figure}

Figure~\ref{Ratio} (solid circles) shows the integrated 
intensity of the 2$D$ peak, $A(2D)$ as a function of the 
integrated intensity of the 2$D'$ peak, $A(2D')$, both 
normalized to the integrated intensity of the $G$ peak, 
$A(G)$, calculated from spectra of Fig.~\ref{Ramanserie}. 
This dependence, besides showing the common decrease of 
the intensity values, as expected from 
Fig.~\ref{Ramanserie}, unveils a constant $A(2D)/A(2D')$ 
ratio, as they closely follow a straight line. The 
decrease in intensity is a direct consequence of the 
increase of $\gamma_{e-e}$ with doping [cf. 
Eqs.~(\ref{subeq:2}) and (\ref{subeq:1})]. On the 
contrary, the constant $A(2D)/A(2D')$ ratio indicates a 
weaker dependence of $\gamma_{e-ph}$ on doping. This point 
becomes clear taking the ratio of Eqs.~(\ref{subeq:2}) and 
(\ref{subeq:1}), which is found to be proportional to the 
square of the emission rates ratio of $E_g$ and $A_1$ 
phonons,
\begin{eqnarray}
  A(2D)/A(2D') = 2(\gamma_{K}/ \gamma_{\Gamma})^2.
  \label{intratio}
\end{eqnarray}
The linear fit to the single-layer data in 
Fig.~\ref{Ratio} gives a value of the slope of 26$\pm$ 
3$\%$ and, from Eq.~(\ref{intratio}), we then obtain the 
ratio of phonon emission rates, $\gamma_{K}/ 
\gamma_{\Gamma} = \ 3.6$. This ratio is related to the 
adimensional EPC parameters $\lambda_{\Gamma}$ and 
$\lambda_{K}$, as defined in  Ref. \onlinecite{Basko08}, 
according to
\begin{eqnarray}
  \frac{\gamma_{K}}{\gamma_{\Gamma}} =    
\frac{\omega_{out,K}}{\omega_{out,\Gamma}}\frac{\lambda_{K}
}{\lambda_{\Gamma}},
\label{rel1}
\end{eqnarray}
where $\omega_{out,\Gamma}$ and $\omega_{out,K}$ are the 
frequecies of the emitted photons in the respective Raman 
processes involving phonons at $\bm{\Gamma}$ and 
$\textbf{K}$ point. In order to compare with available 
calculated and experimental values, we relate the EPCs 
parameters to the square of the $e$-$ph$ interaction 
matrix elements averaged on the Fermi surface, 
$\left\langle D^{2}_{\Gamma}\right\rangle _{F}$ and 
$\left\langle D^{2}_{K}\right\rangle _{F}$, as 
follows~\cite{Basko08,calandra:205411}
\begin{eqnarray}
  \frac{\lambda_{K}}{\lambda_{\Gamma}} = 
  \frac{\omega_{\Gamma}}{\omega_{K}}\frac{\left\langle 
  D^{2}_{K}\right\rangle _{F}}{2\!\left\langle 
  D^{2}_{\Gamma}\right\rangle _{F}},
\label{rel2}
\end{eqnarray}
where $\omega_{\Gamma}$ and $\omega_{K}$ are the 
frequencies of the phonons at $\bm{\Gamma}$ and 
$\textbf{K}$ point, respectively.
%
%**********************************************************
% Figure 4 Intensity areas data for monolayer, bilayer and multylayer
%
%**********************************************************
%
While the value of $\left\langle D^{2}_{K}\right\rangle 
_{F}$ of graphene and related graphitic structures has 
been controversial, due to the lack of reliable 
experimental data of the phonon dispersion around the 
$\textbf{K}$ point, until recently,~\cite{Grueneis:085423} 
the value of $\left\langle D^{2}_{\Gamma}\right\rangle 
_{F}$ was obtained~\cite{lazzeri:155426} by relating it to 
the measured dispersion of the $E_g$ mode near 
$\bm{\Gamma}$.~\cite{maultzsch:075501} Then taking 
$\left\langle D^{2}_{\Gamma}\right\rangle _{F} = 39 
\;(eV/\AA)^{2}$, from Eqs.~(\ref{rel1}) and (\ref{rel2}) 
we obtain $\left\langle D^{2}_{K}\right\rangle _{F} = 205 
\; (eV/ \AA)^{2}$. The latter is close to the EPC value 
found experimentally from the phonon dispersion around the 
$\textbf{K}$ point of graphite,\cite{Grueneis:085423} and 
to the computed EPC values of graphene and graphite, when 
nonlocal exchange-correlation effects are 
included.~\cite{lazzeri:081406} Actually, $\left\langle 
D^{2}_{K}\right\rangle _{F}$ of graphene and graphite are 
expected to differ slightly due to a large screening 
effect of the exchange interaction in the latter. We 
measured the $A(2D)/A(2D')$ ratio in HOPG and found a 
value about 20 \% lower than in graphene (cf. inset to 
Fig.~\ref{Ratio}). For samples with thicknesses of around 
twenty graphene layers the values of the $A(2D)/A(2D')$ 
ratio are midway between graphene and graphite. The value 
of $\left\langle D^{2}_{K}\right\rangle _{F}$ is then 
found to be in the range of 205--188 $(eV/ \AA)^{2}$.

Finally, concerning the doping dependence of EPC 
parameters in graphene, it has been calculated that 
$\left\langle D^{2}_{K}\right\rangle _{F}$ reduces by 
$\approx 16 \%$ for a variation of $\Delta p = 1.9 \cdot 
10^{13}$ cm$^{-2}$, while $\left\langle 
D^{2}_{\Gamma}\right\rangle _{F}$ remains 
unaffected.~\cite{attaccalite10} Taking into account that 
our estimated $\Delta p_{tot}$ is about four times 
smaller, the expected decrease of 
$\left\langle D^{2}_{K}\right\rangle _{F}$ stays within 
the experimental error. 
\subsection{Electron-electron coupling and number of graphene layers}
In Fig.~\ref{Ratio}, we included data of Raman 
measurements from bilayer and ten-layer graphene deposited 
on the same SiO$_2$/Si wafer as the monolayer sample, 
solid squares and triangles, respectively. The 
experimental Raman intensities show a good correlation 
with the monolayer data, as they closely follow the linear 
relation with slope value of 26 (thick line in 
Fig.~\ref{Ratio}). A significant feature is the decrease 
of the intensity rate change, given by the difference 
between intensities in successive treatment loops, which 
is seen as a higher concentration of data points in Fig. 
\ref{Ratio}, when comparing single-layer (SL), bilayer 
(BL) and ten-layer (TL) samples. The dot-dashed line 
arrows in Fig.~\ref{Ratio} are plotted to illustrate the 
data contraction with increasing number of layers, which 
is found to be related by the ratio 
$\Delta A(2D)_{SL} \approx 2\, \Delta A(2D)_{BL} \approx 
10 \, \Delta A(2D)_{TL}$, within an error of 
approximatively 10 \%. Since, as discussed previously, the 
decrease of $A(2D)$ and $A(2D')$, that follows ozone 
treatment, is related to the increase of hole 
concentration, the result of Fig. \ref{Ratio} suggests 
that the rate of decrease of the intensity, $\Delta A(2D)$ 
[$\Delta A(2D')$], with doping becomes smaller by a factor 
inversely proportional to the number of graphene layers. 

In order to understand the relation $\Delta A(2D) \propto 
N^{-1}$ [$\Delta A(2D') \propto N^{-1}$], we first recall 
that a Bernal stack with $N$ layers, $N$ even [$N$ odd], 
has $N/2$ [$(N+1)/2$] electron-like and $N/2$ [$(N+1)/2$] 
hole-like subbands almost touching at the $\textbf {K}$ 
point, with a band overlap smaller than 41 meV. Additional 
$N/2$ [$(N-1)/2$] electron-like and $N/2$ [$(N-1)/2$] 
hole-like outer subbands appear with decreasing energy 
separation with layer number, reaching a maximum value of 
0.4 eV in bilayer graphene. The outermost subband is found 
in the range from 0.4 eV in bilayer to $\approx$ 0.8 eV 
for 20 graphene layers.~\cite{PhysRevB.74.075404} As the 
number of layers increases, the number of $e$-$ph$ 
processes that contributes to the 2$D$ (2$D'$) integrated 
intensity also grows  and it is determined by the 
selection rules for optical excitations and for electron 
scattering by phonons.~\cite{Malard:125426} Since the 
integrated intensities contain the weighted probability of 
the different processes involving phonons with close 
wavevector, the A(2$D'$)/A($2D$) ratio only depends on the 
EPC parameters which, up to ten layers, remain constant as 
inferred from the lineal relation of the data in 
Fig.~\ref{Ratio}. The $\gamma_{e-e}$ scattering rate, on 
the other hand, experiences an increase as the 
intersubband $e$-$e$ collisions are allowed, following the 
appearance of more subbands with increasing number of 
layers. Although for the E$_F$ range studied in the 
present work the parabolicity of the energy subbands 
should be taken into account, for illustrative purposes 
the high doping case is considered where the energy 
dispersion of the subbands can be taken as 
linear.~\cite{Ando07,Das09} Then, in the approximation of 
small momentum transfer, we can generalize Eq. 
(\ref{eerate}) to $N$ layers as, $\gamma_{e-e} = N 
\left|E_{F} \right| f$. Considering linear dispersive 
subbands, the carrier concentration, $p$, is given by $p = 
N \mu E^{2}_{F}$. For highly doped samples, the 
contribution from $e$-$ph$ scattering to the total 
scattering rate can be neglected compared to the much 
bigger $e$-$e$ scattering component. Thus, the integrated 
intensity for multilayer graphene can be written as:
\begin{equation}
	A(2D)_{NL} \approx C \frac{\gamma^{2}_{K}}{N^{2} 
	f^{2}\frac{p}{N}},
	\label{IntN}
\end{equation}
and its rate of decrease upon the change of the carrier 
concentration, $\Delta p$, is
\begin{equation}
	\frac{\Delta A(2D)}{\Delta p} =  - C  
	\frac{\gamma^{2}_{K}}{N f^{2}p^{2}},
	\label{deltaIntN}
\end{equation}
which accounts for the contraction of the data in 
Fig.~\ref{Ratio} with increasing number of layers. 

Based on Eq. \ref{deltaIntN}, the 
approximate dependence of the experimental two-phonon 
Raman intensity with the inverse of $N$ proves that the 
amount of initial, $p$, and 
transferred, $\Delta p$, charge concentration in the 
different samples, SLG, BLG and TLG, ought to be similar. 
Taking into account that all samples 
were placed on the same wafer and exposed to the same 
treatments, it is reasonable to assume that the extent of 
unintentional or background doping and 
adsorbate coverage was comparable among them. The 
dependence of the two-phonon integrated intensity on the 
number of layers is two-fold, as seen from 
Eq.~(\ref{IntN}). First, the charge concentration is 
distributed among the subbands, the number of which 
increase with the number of layers, leading to a decrease 
of the probability of $e$-$e$ collision. Second, with 
increasing number of subbands the number of allowed 
$e$-$e$ processes also increases and, with it, the 
probability of $e$-$e$ collision. Nevertheless, the 
overall effect is a decrease of the two-phonon Raman 
intensity because there is the contribution in the 
scattering rate from the simultaneously excited electron 
and hole, as the square in Eq. (\ref{subeq:2}) reflects. 
\section{Conclusions}
In conclusion, we have shown that the carrier 
concentration in graphene gradually increases with 
sequential ozone short-exposure cycles while preserving 
its crystallinity. The blue-shift of the $G$ and $2D$ peak 
frequencies is evidence of $p$ doping of the samples. In 
contrast to the $G$ peak intensity, which is found to 
remain constant, the $2D$ and $2D'$ peak intensities 
decrease with increasing number of ozone exposure cycles, 
i.e., with increasing doping. This effect reflects the 
responsiveness of the two-phonon Raman intensity to the 
dynamics of photoexcited $e$-$h$ pairs and, in particular, 
to the contribution of the $e$-$e$ scattering. We used 
this dependence to extract the EPC of phonons near the 
$\textbf{K}$ point and found a close agreement with 
previous experimental and theoretical values. We also 
demonstrated an inverse dependence of the rate of decrease 
of the intensities upon doping on the graphene number of 
layers, reflecting the increased probability of $e$-$e$ 
scattering with increasing number of layers.

%
%\acknowledgements
We are grateful to M. Lira-Cant\'u for the use of the 
ozone chamber. Support from the Spanish Ministry of 
Science and Innovation (projects FIS2008-06830 and 
FIS2009-10150) and the EU project NANOPACK (project No 
216176) is gratefully acknowledged.
%
%\nocite{*}
%
\bibliography{graphene_raman01bib}% Produces the 
%bibliography via BibTeX.
%
\end{document}